\documentclass[review,number,sort&compress]{elsarticle}
\usepackage[latin9]{inputenc}
\usepackage{geometry}
\usepackage{mathrsfs}
\usepackage{amsmath}
\usepackage{graphicx}
\usepackage[unicode=true,
 bookmarks=false,
 breaklinks=false,pdfborder={0 0 1},backref=false,colorlinks=false]
 {hyperref}

\makeatletter

\journal{Nuclear Instruments and Methods in Physics Research A}

\usepackage{subfig}

\makeatother

\begin{document}
\begin{frontmatter}

\title{Statistical Significance of CP Violation in Long Baseline Neutrino
Experiments }

\author{Walter Toki{*}, Thomas W. Campbell, Erez Reinherz-Aronis}

\address{Department of Physics, Colorado State University, Fort Collins, CO.
80523, USA}

\cortext[mycorrespondingauthor]{corresponding author, walter.toki@colostate.edu}
\begin{abstract}
The p-value or statistical significance of a CP conservation null
hypothesis test is determined from counting electron neutrino and
antineutrino appearance oscillation events. The statistical estimates
include cases with background events and different data sample sizes,
graphical plots to interpret results and methods to combine p-values
from different experiments. These estimates are useful for optimizing
the search for CP violation with different amounts of neutrino and
antineutrino beam running, comparing results from different experiments
and for simple cross checks of more elaborate statistical estimates
that use likelihood fitting of neutrino parameters.
\end{abstract}
\end{frontmatter}


\section{Introduction}

The search for charge-parity (CP) symmetry violation in neutrino interactions\citep{key-1}
is a major effort in the current\citep{key-5-1} and future\citep{key-3}
particle physics program. A violation of this fundamental symmetry
may be related to the matter-antimatter imbalance observed in the
universe and testing this symmetry is now a fundamental physics question.
This search can be performed using long baseline neutrino experiments
by measuring the neutrino and antineutrino appearance oscillation
probabilities, $P\left(\nu_{\mu}\rightarrow\nu_{e}\right)$ and $P\left(\overline{\nu}_{\mu}\rightarrow\overline{\nu}_{e}\right)$,
respectively. The conservation of the CP symmetry predicts these probabilities
are equal in vacuum,
\begin{equation}
P\left(\nu_{\mu}\rightarrow\nu_{e}\right)=P\left(\overline{\nu}_{\mu}\rightarrow\overline{\nu}_{e}\right)
\end{equation}

These probability values depend\citep{key-1} on neutrino mixing angles,
mass splittings, CP phase and $E/L$, where $E$ is the neutrino or
the antineutrino energy and $L$ is the oscillation distance between
the neutrino origin and its measurement position. CP conservation
(CPC) predicts while travelling in vacuum that the neutrino probability
and the antineutrino probability are equal. If these probabilities
are not equal, then there is CP violation (CPV) in the neutrino sector.
In the case of neutrinos travelling in dense matter over very long
distances, these probabilities can differ due to the MSW effect\citep{key-msw}.
The purpose of this paper is to examine various p-value hypothesis
tests of CP conservation (called the null hypothesis) in long baseline
neutrino test measurements

The experimental setup typically has a proton beam striking a target
which produces positive and negative mesons that are selectively focused
into a long decay pipe to produce either predominately $\nu_{\mu}$
and $\overline{\nu}_{\mu}$ beams, that are measured in a near neutrino
detector and which continue to travel a long baseline distance to
be observed in a far neutrino detector. As the $\nu_{\mu}$ and $\overline{\nu}_{\mu}$
beams travel to the far detector a fraction may under go oscillations
into $\nu_{e}$ and $\overline{\nu}_{e}$ events, respectively, which
are detected in the far detector. The neutrino and antineutrino rates
are determined by measurements of the following inclusive interactions,
\begin{equation}
\begin{array}{ccc}
\nu_{\mu}+N & \rightarrow & \mu^{-}+X\\
\overline{\nu}_{\mu}+N & \rightarrow & \mu^{+}+X\\
\nu_{e}+N & \rightarrow & e^{-}+X\\
\overline{\nu}_{e}+N & \rightarrow & e^{+}+X
\end{array}
\end{equation}

The direct test of CP conservation uses four sample populations which
are the observed numbers of the unoscillated $\nu_{\mu}$ and $\overline{\nu}_{\mu}$
events at the near detector and the $\nu_{e}$ and $\overline{\nu}_{e}$
appearance events at the far detector. The main statistical errors
will come from the limited sample size of the appearance events. A
simple test statistic\citep{key-skellam} can be formed that is the
difference of the two numbers of appearance events with an assumed
known value of the CPC oscillation probabilities. Measurements of
this difference can be used to calculate the p-value probability that
a null hypothesis (CP conservation) could obtain at least this difference
or more extreme values. In the next two sections simple examples are
presented and then neutrino and antineutrino measurements are introduced.
This is followed by sections describing specific neutrino cases. 

\section{Examples of null hypothesis tests using differences of two Poisson
distributions}

Several Poisson type test statististics\citep{key-5-2} have been
used extensively in other research areas for p-value calculations.
An illustrative example is finding the probability that two soccer
teams, that are rated to score the same goals per game, would have
a game with a difference of $\Delta$ goals or larger. Suppose teams
A and B are expected to both score on average 2.5 goals per game.
A test statistic, called a Skellam distribution\citep{key-skellam},
of the difference in their scores forms a random integer variable
that is the difference of two Poissons. The probability $P^{SK}$
as a function of the integer difference $\Delta$ is the convolution
of the two Poissons with averages $\lambda_{A}=\lambda_{B}=2.5$,
\begin{equation}
P^{SK}\left(\Delta\right)=\sum_{n=-\infty}^{\infty}\frac{e^{-\lambda_{A}}\lambda_{A}^{\left(\varDelta+n\right)}}{\left(\varDelta+n\right)!}\frac{e^{-\lambda_{B}}\lambda_{B}^{n}}{n!}
\end{equation}
The Skellam distribution is a 1-dimensional test statistic that is
the probability $P^{SK}\left(\Delta\right)$ versus $\Delta=n_{A}-n_{B}$.
This is plotted in Fig. 1 (a).

\begin{figure}
\hfill{}\subfloat[]{\includegraphics[scale=0.2]{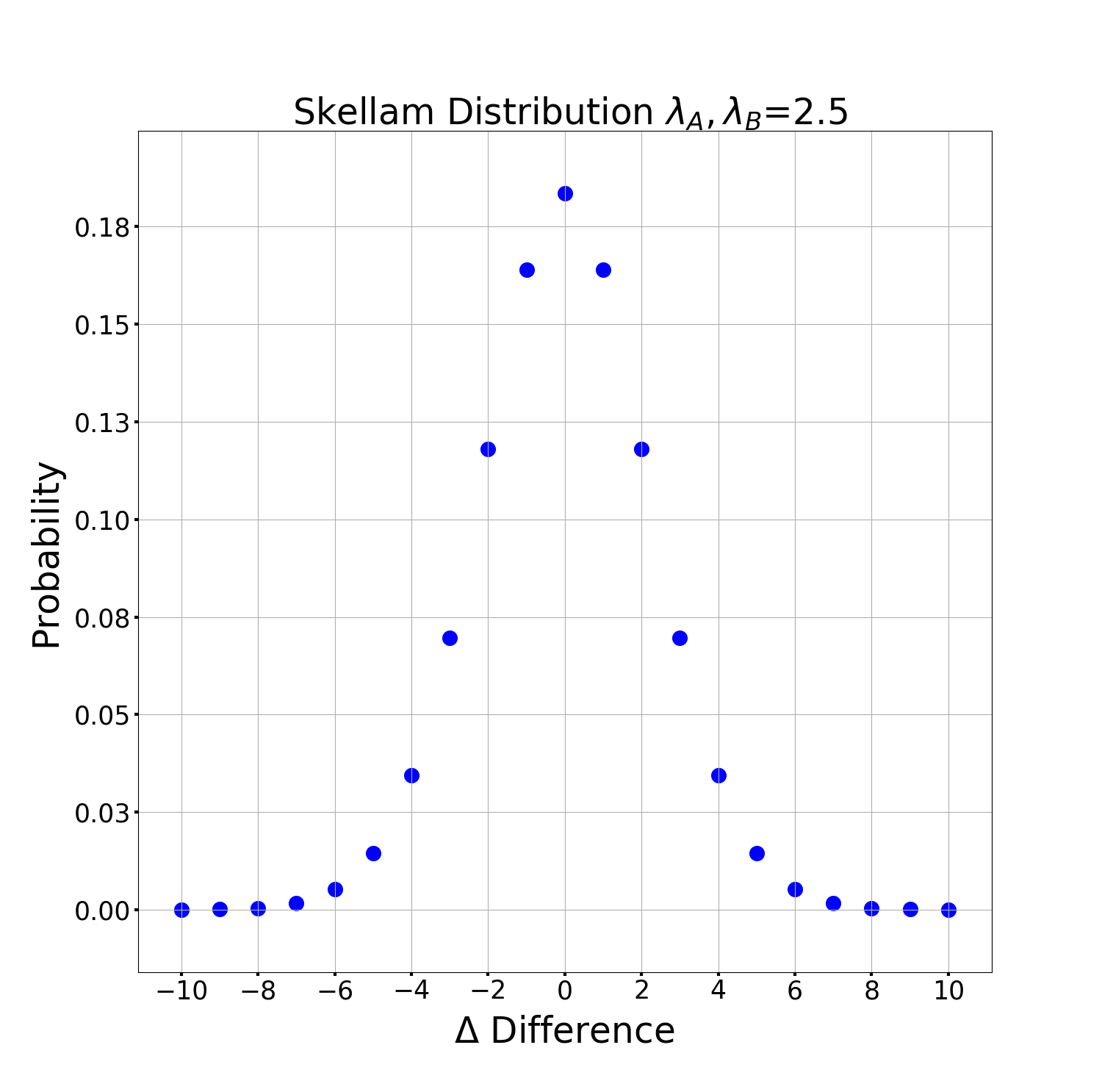}} \subfloat[]{\includegraphics[scale=0.53]{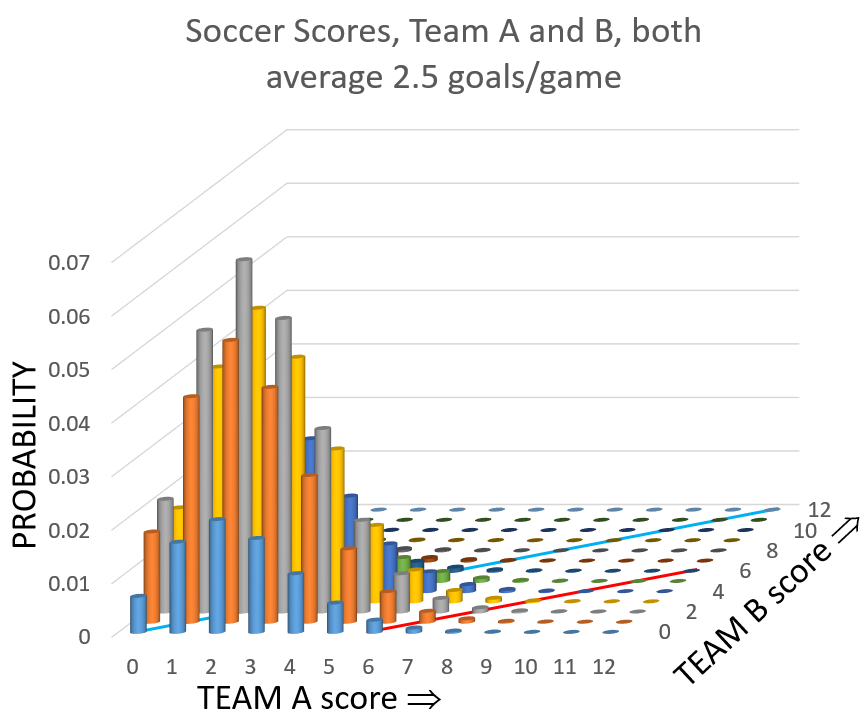}}\hfill{}

Fig. 1. The Skellam distribution, shown in (a), of the difference
of scores between soccer team A and B assuming each team has an average
of 2.5 goals per game. Note if team A and B has a difference of 6
goals, then this corresponds to a probability of 0.00534 and the difference
of obtaining 6 or more goals is 0.00774 which is the sum of probabilities
for $\Delta\geq6$ . In (b) is the Poisson probability Lego plot of
team A versus team B scores. The blue diagonal lines, $N_{A}=N_{B}$,
corresponds to both teams getting the same score which is equivalent
to the null hypothesis of team A and B having equal average scores.
The red diagonal line corresponds to a difference boundary, $N_{A}=N_{B}+6$,
with team A getting 6 goals more than team B. The sum of the probabilities
of all scores on and to the right of the red line correspond to the
p-value of a particular team A getting 6 or more goals than team B.
Note that the diagonal lines are parallel with slope 1 and separated
by 6 goals.
\end{figure}

Suppose in a game, the teams had $N_{A}$=7 and $N_{B}$=1 goals.
This particular game (or test measurement) with a difference of 6
goals has a probability of $P^{SK}\left(\Delta=6\right)=5.34\times10^{-3}$.
The probability of obtaining difference of 6 $or$ $more$ goals is,

\begin{equation}
\sum_{N_{A}=0}^{\infty}\sum_{N_{B}=0}^{\infty}\theta\left(N_{A}-N{}_{B}-\Delta+\epsilon\right)\frac{e^{-\lambda_{A}}\lambda_{A}^{N_{A}}}{N_{A}!}\frac{e^{-\lambda_{A}}\lambda_{B}^{N_{B}}}{N_{B}!}
\end{equation}
which sums to $7.74\times10^{-3}$ assuming $\lambda_{A}=\lambda_{B}=2.5$.
This corresponds to the p-value which is the probability the null
hypothesis, which assumes the same average $2.5$, can reach this
extreme value or higher. The joint Poisson probabilities as a function
of team A and B scores are displayed in Fig. 1(b). The blue diagonal
line, $N_{A}=N{}_{B}$, corresponds to both teams having equal score
probabilities. The red line, $N_{A}=N_{B}+6$, is the boundary, parallel
to the blue line, that represents the difference of team A having
six goals more than team B. Each bin of the Skellam distribution sums
the probabilities in Fig. 1(b), in between two diagonal lines that
are separated by a perpendicular distance of $1/\sqrt{2}$.

The test measurement p-value of the null hypothesis (both teams had
the same average score 2.5 goals) is the probability that a difference
of 6 or higher goals occurs and in this example the p-value is 0.0024.
This corresponds to a 3 standard deviation test of the null hypothesis
that the teams had the same average number of goals per game. 

Another example is a test if two radioactive samples have the same
decay rates. Suppose there are two samples of radioactive nuclei type
A and B with initially $N_{A}$ and $N_{B}$ nuclei, respectively,
which are equal. Then suppose there are measurements of $n_{A}$ and
$n_{B}$ decays in a time t. The null hypothesis is if these two samples
have the same decay rate $\varGamma$. This can be tested by calculating
the p-value of the null hypothesis where $n_{A}=N_{A}\left(1-e^{-\Gamma t}\right)$
and $n_{B}=N_{B}\left(1-e^{-\Gamma t}\right)$. In this example, suppose
a measurement of the samples yielded $n_{A}$ and $n_{B}$. The p-value
that the radioactive samples are the same is calcuated with Eq. (6)
where $\lambda_{A}=n_{A}$= $\lambda_{B}=n_{B}$ and $\Delta=n_{A}-n_{B}$.

\section{Neutrino and Antineutrino Measurements}

The test of CP violation in a long baseline neutrino experiment is
analogous to testing if two radioactive nuclei samples have the same
decay rates. The neutrino far detector counts the number of observed
neutrino events $N_{e}$ found in a $\nu_{\mu}$ beam and the number
of observed antineutrino events $\overline{N}_{e}$ found in a $\overline{\nu}_{\mu}$
beam and the difference is considered. The appearance probabilities
in terms of observed events are, 

\begin{equation}
P\left(\nu_{\mu}\rightarrow\nu_{e}\right)=\frac{N_{e}}{N_{\mu}}\text{ and }P\left(\overline{\nu}_{\mu}\rightarrow\overline{\nu}_{e}\right)=\frac{\overline{N}_{e}}{\overline{N}_{\mu}}\label{eq:antinu-prob}
\end{equation}
These number of events are averaged over $\left\langle E_{\nu}/L\right\rangle $
and $\left\langle E_{\overline{\nu}}/L\right\rangle $ where the neutrino
and antineutrino energies should be identical. A test of the CPC hypothesis
is performed by determining the p-value of the null hypothesis. Given
experimental measurements of the $N_{e}$ and $\overline{N}_{e}$
and unoscillated $N_{\mu}$ and $\overline{N}_{\mu}$, the above probabilities
can be obtained. The null hypothesis of CPC can also be formed using
a test statistic such as the asymmetry $A$, 

\begin{equation}
A=\frac{P\left(\nu_{\mu}\rightarrow\nu_{e}\right)-P\left(\overline{\nu}_{\mu}\rightarrow\overline{\nu}_{e}\right)}{P\left(\nu_{\mu}\rightarrow\nu_{e}\right)+P\left(\overline{\nu}_{\mu}\rightarrow\overline{\nu}_{e}\right)}\label{eq:asymmetry}
\end{equation}
or the difference in the appearance rates, denoted as $\Delta$,

\begin{equation}
\Delta=N_{e}-\overline{N}_{e}\label{eq:Delta}
\end{equation}
This assumes the unoscillated muon neutrino and antineutrino events
samples at the far detector are the same, $N_{\mu}=\overline{N}_{\mu}$.
This tests how unequal the probabilities are since, 
\begin{equation}
P\left(\nu_{\mu}\rightarrow\nu_{e}\right)-P\left(\overline{\nu}_{\mu}\rightarrow\overline{\nu}_{e}\right)=\frac{N_{e}-\overline{N}_{e}}{N_{\mu}}
\end{equation}
and it is seen that the p-value of a measurement difference test statistic
$\Delta$ or more extreme is equivalent to measuring the corresponding
neutrino-antineutrino probability difference in Eq. (6) or larger.
If there are unequal sample sizes where $R=N_{\mu}/\overline{N}_{\mu}\neq1$,
then the above probability difference, $A$ and $\Delta$ become,

\begin{equation}
P\left(\nu_{\mu}\rightarrow\nu_{e}\right)-P\left(\overline{\nu}_{\mu}\rightarrow\overline{\nu}_{e}\right)=\frac{N_{e}}{N_{\mu}}-\frac{\overline{N}_{e}}{\overline{N}_{\mu}}=\frac{1}{N_{\mu}}\left(N_{e}-R\overline{N}_{e}\right)
\end{equation}

\begin{equation}
A=\frac{N_{e}-R\overline{N}_{e}}{N_{e}+R\overline{N}_{e}}
\end{equation}
\begin{equation}
\Delta=N_{e}-R\overline{N}_{e}
\end{equation}
The $N_{e}$ and $\overline{N}_{e}$ are number of observed events
averaged over neutrino energy. Again the p-value of the test statistic
in Eq. (8) is equivalent to the p-value of the neutrino-antineutrino
probability difference in Eq. (6). If the detector and the reconstruction
efficiencies, the $\nu_{e}$ cross section, the $\overline{\nu}$$_{e}$
cross section and the number of observed $N_{\mu}$ and $\overline{N}_{\mu}$
are included, the ratio $R$ becomes,
\begin{equation}
R=\left(\frac{\overline{\epsilon}_{e}}{\epsilon_{e}}\right)\frac{N_{\mu}/\epsilon_{\mu}}{\overline{N}_{\mu}/\overline{\epsilon}_{\mu}}
\end{equation}
where $\epsilon_{e}$ and $\overline{\epsilon}_{e}$ are the efficiencies
to reconstruct far detector events and $\epsilon_{\mu}$ and $\overline{\epsilon}_{\mu}$
are the efficiences to reconstruct the near detector events. Note
a Poisson probability plotted as a function of $N_{e}$ versus $\overline{N}_{e}$,
is analogous to Fig. 1(b), where the line representing equal rates,
$0=N_{e}-R\overline{N}_{e}$, and the difference boundary $\Delta=N_{e}-R\overline{N}_{e}$
will be parallel to each other with slope $R$. The simple methods
described here, should be useful to estimate p-value probabilities
of the null CPV hypothesis. They require far detector measurements
of the $N_{e}$ events in a neutrino beam, the $\overline{N}_{e}$
events in an antineutrino beam and the ratio of unoscillated rates
of $\overline{N}_{\mu}/N_{\mu}$ at the far detector. If the long
baseline experiment $E/L$, has muon neutrino disappearance that oscillates
into zero events at the far detector, then a near detector will be
essential to measure this ratio $\overline{N}_{\mu}/N_{\mu}$ and
extrapolate the unoscillated rate ratio to the far detector. More
complex methods to determine CPV include fitting the $E/L$ distributions
to determine the Pontecorvo, Maki, Nakagawa and Sakata (PMNS) neutrino
parameters\citep{key-1}, but due to the expected small statistics
from neutrino long baseline neutrino experiments in the near term,
the methods presented here should be adequate to predict p-value tests
of CP violation. The difference test statistic $\Delta$ will not
be affected by uncertainties of the PMNS neutrino parameters that
include the mixing angles, CP phase and the mass hierarchy, however
the magnitude of the probability depends on these PMNS parameters
and this can affect the resulting p-value using a Skellam distribution.
The aim in this paper is to test for CPV in Eqn. 1, independent of
the values of the PMNS parameters.

In the following sections, different cases are presented with pedagogical
examples. The material is presented in order of complexity and include
cases; (I) equal data samples with no background, (II) equal data
samples with background, (III) different data sample sizes, (IV) different
sample sizes with backgrounds and (V), where the oscillation probabilies
are assumed to be unknown. Finally a method to combine p-values from
2 different experiments is presented. Simple calculations and formulae
are provided so the reader can replicate the results. In the appendix,
the general case of an integrated flux measurement, cross sections,
detection efficiencies and how their affects are included in the $R$
factor is described.

\section{Case I. No Background and Equal Data Samples}

The simplied case with no background and equal amounts of neutrino
and antineutrino data is presented here. It is assumed, the reconstruction
efficiency of the electron and muon neutrino interactions are the
same and the uncertainties due to the numbers of $\nu_{\mu}$ and
$\overline{\nu}_{\mu}$ are small compared to the uncertainties of
the number of $\nu_{e}$ and $\overline{\nu}_{e}$ to simplify the
equations. The simpliest test statistic is the difference in the number
$\nu_{e}$ and $\overline{\nu}_{e}$ appearance events. 

For our neutrino example, suppose at the far detector, the sample
of the number of unoscillated muon neutrino and antineutrinos that
could be reconstructed are the same, 
\begin{equation}
N_{\mu}=\overline{N}_{\mu}
\end{equation}
If CPC is true, the number of expected or average number of appearance
events for neutrinos $\lambda$ and for antineutrinos $\overline{\lambda}$
should be equal. \ The probability of a particular observation of
$N_{e}$ and $\overline{N}_{e}$ will be,
\begin{equation}
\frac{e^{-\lambda}\lambda^{N_{e}}}{N_{e}!}\frac{e^{-\overline{\lambda}}\overline{\lambda}^{\overline{N}_{e}}}{\overline{N}_{e}!}
\end{equation}
The Poisson probability is denoted as script $\mathscr{P}\left(N,\lambda\right)$
, where the double sum is always normalized, 
\begin{equation}
\sum_{N_{e}=0}^{\infty}\sum_{\overline{N}_{e}=0}^{\infty}\mathscr{P}\left(N_{e},\lambda\right)\mathscr{P}\left(\overline{N}_{e},\overline{\lambda}\right)=\sum_{N_{e}=0}^{\infty}\mathscr{P}\left(N_{e},\lambda\right)\sum_{\overline{N}_{e}=0}^{\infty}\mathscr{P}\left(\overline{N}_{e},\overline{\lambda}\right)=1
\end{equation}
The CPC p-value of observing a difference between the neutrino and
antineutrino appearance rates, $N_{e}-\overline{N}_{e}\geq\Delta$
is given as

\begin{equation}
\sum_{N_{e}=0}^{\infty}\sum_{\overline{N}_{e}=0}^{\infty}\theta\left(N_{e}-\overline{N}_{e}-\Delta+\epsilon\right)\mathscr{P}\left(N_{e},\lambda\right)\mathscr{P}\left(\overline{N}_{e},\overline{\lambda}\right)\label{eq:base-eqn}
\end{equation}
where $\theta\left(x\right)$ is the Heaviside or step function and
where $1>\epsilon>0$ is added to ensure that the term where $N_{e}-\overline{N}_{e}=\Delta$
is counted. Note this can be rewritten in terms of modified Bessel
function\citep{key-skellam}. If a difference $\Delta$ is observed,
the above formula is the extreme probability that the CPC hypothesis
must have to produce a difference of $\Delta$ or larger.

Using the PMNS mixing matrix, the mixing angles and a nonzero CP\ phase,
the CPV $\lambda$ and $\overline{\lambda}$ values and the resulting
$\Delta$ can be determined and used to estimate the CPC p-value.
Suppose for the CPC ($\delta_{CP}=0$) scenario these yield, 
\begin{equation}
P\left(\nu_{\mu}\rightarrow\nu_{e}\right)=P\left(\overline{\nu}_{\mu}\rightarrow\overline{\nu}_{e}\right)=0.051
\end{equation}
and for the CPV ($\delta_{CP}=-\pi/2$) scenario,

\begin{equation}
P\left(\nu_{\mu}\rightarrow\nu_{e}\right)=0.065
\end{equation}

\begin{equation}
P\left(\overline{\nu}_{\mu}\rightarrow\overline{\nu}_{e}\right)=0.037
\end{equation}
Then for sample size of $N_{\mu}=\overline{N}_{\mu}=1000,$ the predictions
for CPC scenario, neutrino appearance in vacuum are,
\begin{equation}
N_{e}=\overline{N}_{e}=51
\end{equation}
and for the CPV scenario the predictions are,

\begin{equation}
N_{e}=65
\end{equation}
\begin{equation}
\overline{N}_{e}=37
\end{equation}
The above CPV case predicts a difference value, $\Delta=65-37=28$
and with
\begin{equation}
\sum_{N_{e}=0}^{\infty}\sum_{\overline{N}_{e}=0}^{\infty}\theta\left(N_{e}-\overline{N}_{e}-28+\epsilon\right)\mathscr{P}\left(N_{e},65\right)\mathscr{P}\left(\overline{N}_{e},37\right)=0.59
\end{equation}
This predicts there is a 59\% chance of observing $\Delta$ with 28
or higher in a single measurement of the CPV scenario. The CPC case
in vacuum has $\lambda$ and $\overline{\lambda}$ equal to 51 and
the probability of a null CPC hypothesis is,
\begin{equation}
\sum_{N_{e}=0}^{\infty}\sum_{\overline{N}_{e}=0}^{\infty}\theta\left(N_{e}-\overline{N}_{e}-28+\epsilon\right)\mathscr{P}\left(N_{e},51\right)\mathscr{P}\left(\overline{N}_{e},51\right)=0.00242
\end{equation}
Hence the p-value is 0.242\% and the resulting double Poisson plot
is given in the Fig. 2(a) with a boundary set by $\Delta=28$. The
blue line in Fig. 2(a) is the boundary which is a 45$^{\circ}$ line
that lies on the boundary point at $N_{e}=65$ and $\overline{N}_{e}=37$.
The Skellam distribution of the double Poisson distribution is plotted
in Fig. 3(a). The arrow points to the 28 difference which corresponds
to the 0.00242 probability which is the area to the left of the blue
line. The p-value depends on the value of $\lambda$ and $\overline{\lambda}$.
If $\lambda$ and $\overline{\lambda}$ were both equal to 46 or 56,
then the resulting p-value (for the same $\Delta=28$) changes to
0.151 or 0.357, respectively. Note, if the mass effects are large,
then unequal values of $\lambda$ and $\overline{\lambda}$ should
be used for the null hypothesis.

\begin{figure}
\hfill{}\subfloat[]{\includegraphics[scale=0.28]{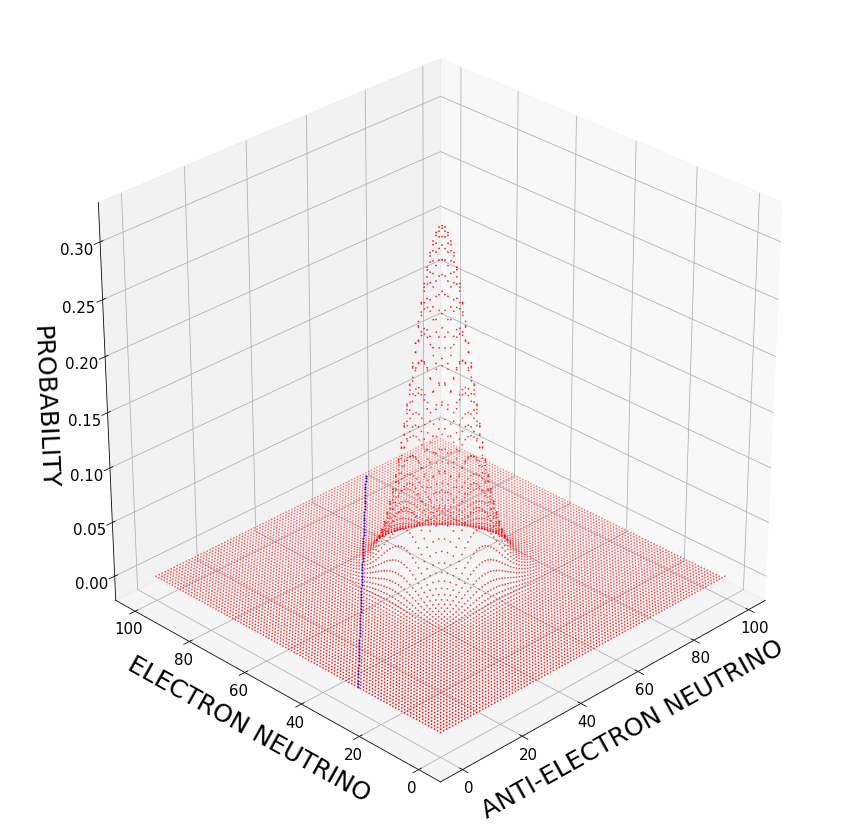}}\subfloat[]{\includegraphics[scale=0.28]{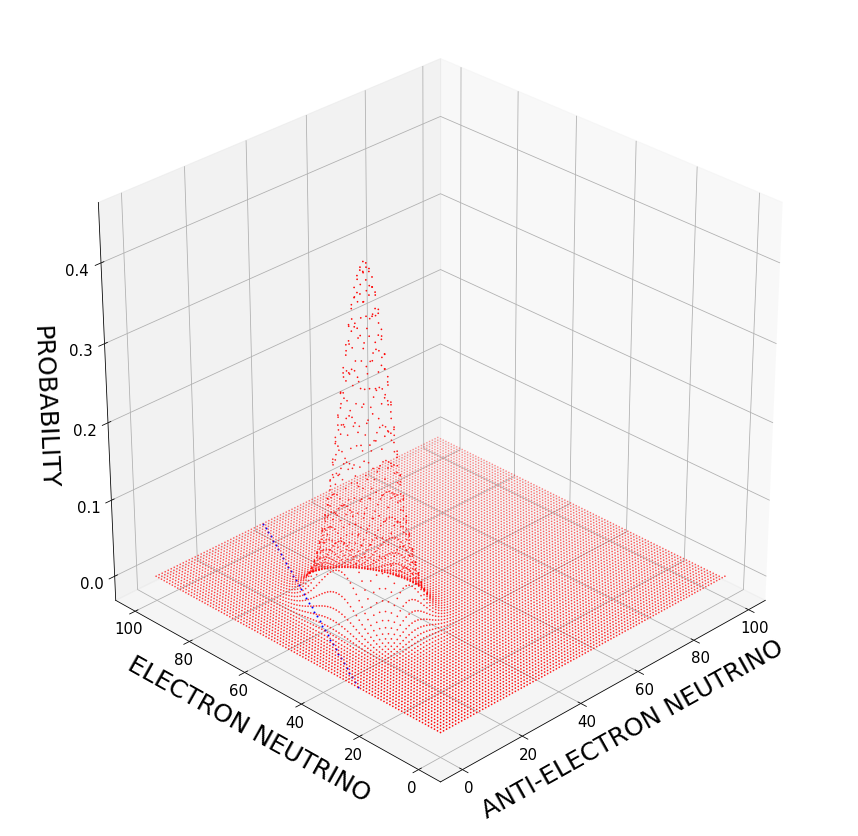}}\hfill{}

\hfill{}\subfloat[]{\includegraphics[scale=0.28]{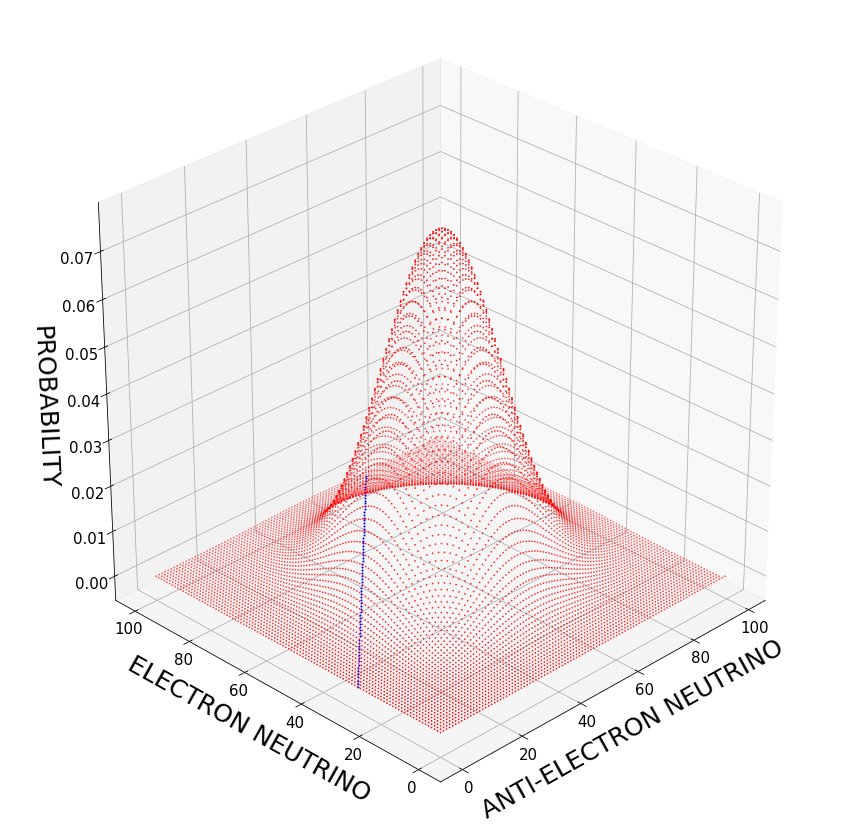}}\subfloat[]{\includegraphics[scale=0.28]{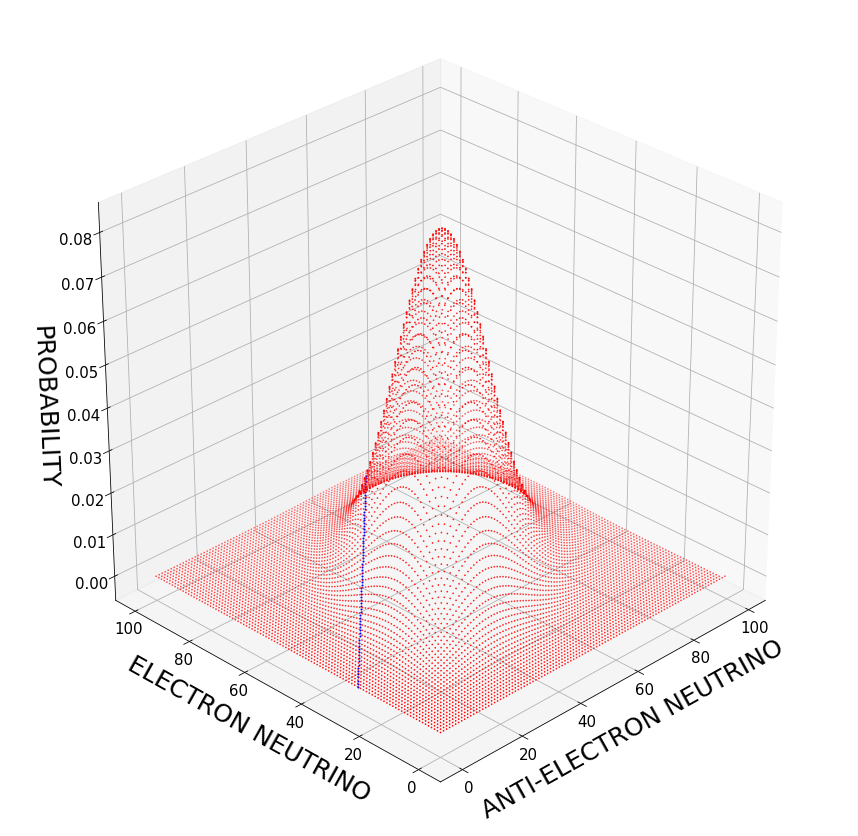}}\hfill{}

Fig. 2. In (a), the red dots are double Poisson distribution of number
of $\nu_{e}$ versus the number of $\overline{\nu}_{e}$ events assuming
that both distributions have an average of 51 events. The vertical
probability axes are in percent. The blue dots represent the difference
$n\left(\nu_{e}\right)-n\left(\overline{\nu}_{e}\right)=28$. The
p-value which is the integrated probabilities of the region to the
left of and including the blue dots is 0.24\%. This represents statistical
p-value of a test measurement of a difference of 28 events. In (b)
The red dots are double Poisson distribution of number of $\nu_{e}$
(with average 51 events) versus the number of $\overline{\nu}_{e}$
events (with average of 28.5 events). The Probabilities are in percent.
The blue dots represent the difference $n\left(\nu_{e}\right)-2\times n\left(\overline{\nu}_{e}\right)=28$.
The p-value or integrated probabilities of the region to the left
of and including the blue dots is 0.9\%. This represents statistical
p-value of a test measurement of a difference of n$\left(\nu_{e}\right)-2\times n\left(\overline{\nu}_{e}\right)\geq28$
events. In (c) and (d), both Poisson distributions are smeared by
25\%, however (d) has a correlation ratio of $\rho=0.5$. Note the
positive correlation makes the distribution more narrow w.r.t. the
blue diagonal and decreases the p-value. This is expected since this
represents a cancellation of correlated neutrino and anti-neutrino
errors. 
\end{figure}

The asymmetry test statistic can be calculated as,

\begin{equation}
A=\frac{\Delta}{N_{e}+\overline{N}_{e}}=\frac{65-37}{65+37}=\allowbreak0.275
\end{equation}
and the CPC probability of an observation is,

\begin{equation}
\sum_{N_{e}=0}^{\infty}\sum_{\overline{N}_{e}=0}^{\infty}\theta\left(\frac{N_{e}-\overline{N}_{e}}{N_{e}+\overline{N}_{e}}-A\right)\mathscr{P}\left(N_{e},51\right)\mathscr{P}\left(\overline{N}_{e},51\right)=0.0024
\end{equation}
Note that the step functions are related,
\[
\theta\left(\frac{N_{e}-\overline{N}_{e}}{N_{e}+\overline{N}_{e}}-A\right)=\theta\left(\frac{N_{e}-\overline{N}_{e}-A\left(N_{e}+\overline{N}_{e}\right)}{N_{e}+\overline{N}_{e}}\right)
\]

\begin{equation}
=\theta\left(N_{e}-\overline{N}_{e}-A\left(N_{e}+\overline{N}_{e}\right)\right)
\end{equation}
so the asymmetry statistic $A$ will have the exactly the same p-value
probability results as the difference of observed events statistic
$\Delta$ where $\Delta=A\left(N_{e}+\overline{N}_{e}\right)$.

Recapping the salient points, if one assumes the CPV ($\delta_{CP}=-\pi/2$)
scenario is true, then there is a 59\% chance of observing a measurement
of at least a difference of 28 events and excluding CP conservation
at the 3 standard deviation level. 

\section{Case II. Different Sample Sizes}

In the previous sections, it was assumed the measurement efficiencies
of $N_{e}$, $N_{\mu}$, $\overline{N}_{e}$ and $\overline{N}_{\mu}$
were the same. Experimentally, the muon and electron neutrino cross
sections should be very similar, however the neutrino and antineutrino
cross sections on the same nuclear targets will be different. Hence
unless the neutrino and antineutrino integrated beam fluxes are adjusted
to equalize the differences, we would expect very different sample
sizes. However, this can be corrected by appropriately modifying the
ratio $R$ defined in Eq. (11). 

In this case, unequal neutrino and antineutrino data sizes, $N_{\mu}\neq\overline{N}_{\mu}$,
are examined. Reconsider case 1 with a antineutrino sample size that
is smaller and corresponds to $\overline{N}_{\mu}=500$. In this case
the CPC average number of observed events is $\overline{\lambda}=\frac{51}{2}=25.5$
and a difference of neutrino and antineutrino events is considered
where the antineutrino rate is scaled up by a factor 2, so the predicted
difference is $65-2\times18.5\geq28$.

\begin{equation}
\sum_{N_{e}=0}^{\infty}\sum_{\overline{N}_{e}=0}^{\infty}\theta\left(N_{e}-2\times\overline{N}_{e}-28+\epsilon\right)\mathscr{P}\left(N_{e},51\right)\mathscr{P}\left(\overline{N}_{e},25.5\right)=0.0090
\end{equation}
This case is displayed in Fig. 2(b). Note that step function is $N_{e}-2\times\overline{N}_{e}-28+\epsilon$
instead of $N_{e}-\overline{N}_{e}-(65-37/2+\epsilon)$. The CPC null
hypothesis is along the line defined by $N_{e}=2\times\overline{N}_{e}$
and the p-value boundary is $N_{e}+28=2\times\overline{N}_{e}$ which
is parallel to the CPC null hypothesis line.

A more extreme case is when $\overline{N}_{\mu}=200$, then p-value
becomes,

\begin{equation}
\sum_{N_{e}=0}^{\infty}\sum_{\overline{N}_{e}=0}^{\infty}\theta\left(N_{e}-5\times\overline{N}_{e}-28+\epsilon\right)\mathscr{P}\left(N_{e},51\right)\mathscr{P}\left(\overline{N}_{e},10.2\right)=0.044
\end{equation}
If the ratio of the neutrino to antineutrino sample sizes is $R$,
then the general expression becomes,
\begin{equation}
\sum_{N_{e}=0}^{\infty}\sum_{\overline{N}_{e}=0}^{\infty}\theta\left(N_{e}-R\overline{N}_{e}-\Delta+\epsilon\right)\mathscr{P}\left(N_{e},\lambda\right)\mathscr{P}\left(\overline{N}_{e},\overline{\lambda}/R\right)
\end{equation}
Note that the diagonal line the represents the boundary for Fig. 2(b),
is has a slope of $R$ and unless $R$ is an integer, the line does
not intersect all the integer points $[N_{e},\overline{N}_{e}]$ in
the x-y plane in Fig. 2(b). In this case, it is not a conventional
Skellam distribution that normally is an $integer$ test statistic
as shown in Fig. 1(a), where the x-axis is in integer steps. If the
ratio $R>1$ and not an integer ($ex.$ $R=\pi$), then it is more
sensible to have a test statistic in units or steps of $R$ such that
the 1-dimensional test statistic represents the summed probabilities
between parallel lines with slope $R$. In this case, it is straight
forward to see that these parallel lines with slope $R$ are separated
by perpendicular distance, $R/\sqrt{R^{2}+1}$, which is consistent
with the Skellam case when $R=1$.

\section{Case III. Backgrounds}

Real particle experiments typically have backgrounds or spurious events
in their signal candidate event sample. This case is examined in this
section. Suppose the antineutrino sample in the previous section has
10 background events. The Poisson probability will be

\begin{equation}
\mathscr{P}\left(N_{e},51\right)\mathscr{P}\left(\overline{N}_{e},51+10\right)
\end{equation}
and the predicted difference in the number of events will be $65-\left(37+10\right)=18$.
\ The probability of a null hypothesis increases to

\begin{eqnarray*}
\sum_{N_{e}=0}^{\infty}\sum_{\overline{N}_{e}=0}^{\infty}\theta\left(N_{e}-\left(\overline{N}_{e}-10\right)-28+\epsilon\right)\mathscr{P}\left(N_{e},51\right)\mathscr{P}\left(\overline{N}_{e},51+10\right) & =\\
\end{eqnarray*}
\begin{equation}
\sum_{N_{e}=0}^{\infty}\sum_{\overline{N}_{e}=0}^{\infty}\theta\left(N_{e}-\overline{N}_{e}-18+\epsilon\right)\mathscr{P}\left(N_{e},51\right)\mathscr{P}\left(\overline{N}_{e},61\right)=0.0035
\end{equation}
And if both neutrino and antineutrino samples had 10 background events,
then the following probability for CPC is,

\begin{equation}
\sum_{N_{e}=0}^{\infty}\sum_{\overline{N}_{e}=0}^{\infty}\theta\left(N_{e}-\overline{N}_{e}-28+\epsilon\right)\mathscr{P}\left(N_{e},61\right)\mathscr{P}\left(\overline{N}_{e},61\right)=0.0050
\end{equation}
The general form the test statistic with backgrounds will be given
as before except the constants are changed by $\lambda\rightarrow\lambda+b$,
$\overline{\lambda}\rightarrow\overline{\lambda}+\overline{b}$, and
$\Delta\rightarrow\Delta+\left(b-\overline{b}\right)$, then,

\begin{equation}
\sum_{N_{e}=0}^{\infty}\sum_{\overline{N}_{e}=0}^{\infty}\theta\left(N_{e}-\overline{N}_{e}-\left(\Delta+b-\overline{b}\right)+\epsilon\right)\mathscr{P}\left(N_{e},\lambda+b\right)\mathscr{P}\left(\overline{N}_{e},\overline{\lambda}+\bar{b}\right)
\end{equation}
If there is uncertainty in $\lambda$ or the background $b$, it is
typically estimated as a Gaussian uncertainty $\sigma$, and the null
hypothesis Poisson probability must be integrated and smeared about
the $x_{0}=\lambda+b$,

\begin{equation}
\frac{1}{\mathscr{N}}\int_{0}^{\infty}e^{-\left(x-x_{0}\right)^{2}/2\sigma^{2}}\mathscr{P}\left(\mathrm{m,x}\right)dx=\frac{1}{N}\int_{0}^{\infty}\frac{e^{-x}x^{m}}{m!}e^{-\left(x-x_{0}\right)^{2}/2\sigma^{2}}dx
\end{equation}
where, $\mathscr{N}=\int_{0}^{\infty}e^{-\left(x-x_{0}\right)^{2}/2\sigma^{2}}dx$.
This integral can be solved numerically, but the definite integral\citep{key-7}
is related to parabolic cylindrical function $D_{n}(z)$ and has a
simple closed form,

\[
\begin{array}{c}
\int_{0}^{\infty}e^{-x}x^{m}e^{-\left(x-x_{0}\right)^{2}/2\sigma^{2}}dx=\\
\\
\end{array}
\]

\begin{equation}
\sigma^{m+1}\Gamma\left(m+1\right)exp\left(-\frac{x_{0}^{2}}{4\sigma^{2}}-\frac{x_{0}}{2}-\frac{\sigma^{2}}{4}\right)D_{-m-1}\left(-\frac{x_{0}}{\sigma}+\sigma\right)
\end{equation}
Using the above, a Poisson convoluted with a Gaussian, $\widetilde{\mathscr{P}}\left(m,x_{0},\sigma\right)$
is defined as
\begin{eqnarray}
 &  & \begin{array}{c}
\widetilde{\mathscr{P}}\left(m,x_{0},\sigma\right)=\frac{1}{N}\int_{0}^{\infty}\frac{e^{-x}x^{m}}{m!}e^{-\left(x-x_{0}\right)^{2}/2\sigma^{2}}dx=\\
=\frac{\sigma^{m+1}\Gamma\left(m+1\right)}{Nm!}exp\left(-\frac{x_{0}^{2}}{4\sigma^{2}}-\frac{x_{0}}{2}-\frac{\sigma^{2}}{4}\right)D_{-m-1}\left(-\frac{x_{0}}{\sigma}+\sigma\right)\begin{array}{c}
\\
\\
\end{array}
\end{array}\label{eq:convolution}
\end{eqnarray}
and then the test statistic becomes,
\begin{equation}
\sum_{N_{e}=0}^{\infty}\sum_{\overline{N}_{e}=0}^{\infty}\theta\left(N_{e}-\overline{N}_{e}-\left(\Delta+b-\overline{b}\right)+\epsilon\right)\widetilde{\mathscr{P}}\left(N_{e},\lambda+b,\sigma\right)\mathscr{\widetilde{P}}\left(\overline{N}_{e},\overline{\lambda}+\overline{b},\overline{\sigma}\right)
\end{equation}
Again, note that the double sum of the smeared Poisson is always normalized, 

\[
\sum_{N_{e}=0}^{\infty}\sum_{\overline{N}_{e}=0}^{\infty}\widetilde{\mathscr{P}}\left(N_{e},\lambda+b,\sigma\right)\mathscr{\widetilde{P}}\left(\overline{N}_{e},\lambda+\overline{b},\overline{\sigma}\right)
\]

\[
=\sum_{N_{e}=0}^{\infty}\widetilde{\mathscr{P}}\left(N_{e},\lambda+b,\sigma\right)\sum_{\overline{N}_{e}=0}^{\infty}\mathscr{\widetilde{P}}\left(\overline{N}_{e},\lambda+\overline{b},\overline{\sigma}\right)
\]

\[
=\frac{1}{\mathscr{N}}\int_{0}^{\infty}\left(\sum_{n=0}^{\infty}\frac{e^{-\lambda}\lambda^{n}}{n!}\right)e^{-\left(\lambda-\lambda_{0}\right)^{2}/2\sigma^{2}}d\lambda
\]

\begin{equation}
\times\frac{1}{\overline{\mathscr{N}}}\int_{0}^{\infty}\left(\sum_{m=0}^{\infty}\frac{e^{-\overline{\lambda}}\overline{\lambda}^{m}}{m!}\right)e^{-\left(\overline{\lambda}-\overline{\lambda}_{0}\right)^{2}/2\overline{\sigma}^{2}}d\overline{\lambda}=1
\end{equation}
If the errors in the neutrino and antineutrino Poisson's are correlated,
$\rho=\frac{\left\langle \sigma\overline{\sigma}\right\rangle }{\sigma\overline{\sigma}}\neq0$,
then a 2 dimensional integral with correlations is formed. The correlated
errors can include the flux errors on the neutrino and antineutrino
beams, the cross section uncertainties and the detector efficiency
uncertainties. The separate integrals becomes a Gaussian bivariate
distribution,

\[
\frac{1}{\overline{\mathscr{N}}}\int_{0}^{\infty}\int_{0}^{\infty}\frac{e^{-x}x^{m}}{m!}\frac{e^{-y}y^{n}}{n!}
\]

\begin{equation}
\times e^{\left(-\frac{1}{1-\rho^{2}}\left[-\frac{1}{2}\left(x-\lambda\right)^{2}/\sigma^{2}-\frac{1}{2}\left(y-\overline{\lambda}\right)^{2}/\overline{\sigma}^{2}+\rho\left(x-\lambda\right)\left(y-\overline{\lambda}\right)/\sigma\overline{\sigma}\right]\right)}dxdy\label{eq:bivariate}
\end{equation}
If the covariance is positive, $\rho>0$, then the p-value will be
reduced since the principle axis of the error elipse that is parallel
to the boundary of the p-value region will be narrower. In Figs. 2(c)-(d),
are the results for 25\% and 25\% with correlation errors of $\rho=0.5,$
respectively. 

\begin{figure}
\hfill{}\subfloat[]{\includegraphics[scale=0.53]{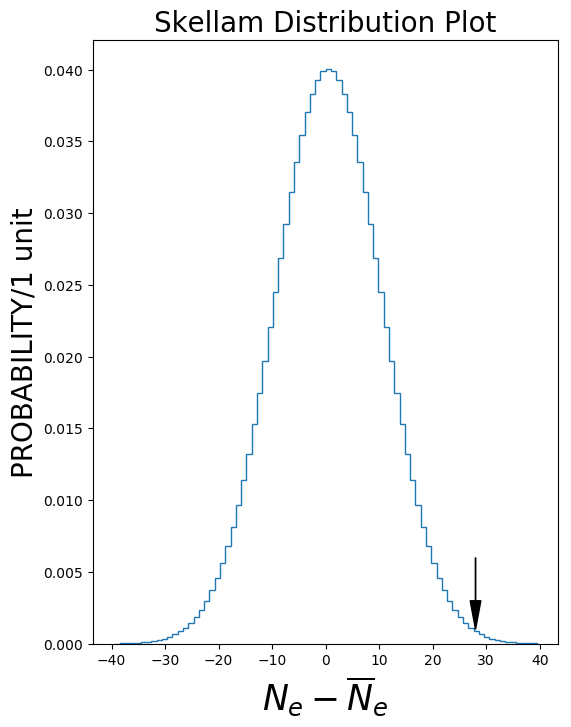}}\subfloat[]{\includegraphics[scale=0.53]{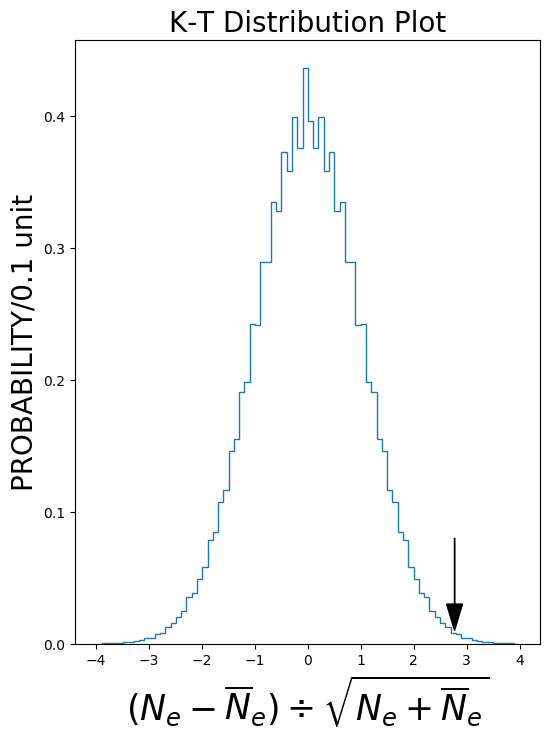}}\hfill{}

Fig. 3. In (a) is the Skellam distribution of the double Poisson distributions
created from Fig. 2(a). The arrow at $N_{e}-\overline{N}_{e}=28$
corresponds to a p-value of 0.0024. The K-T test statistic of the
difference in standard deviations is plotted in (b). The arrow at
$\left(N_{e}-\overline{N}_{e}\right)/\sqrt{N_{e}+\overline{N}_{e}}=2.77$
corresponds to a p-value of 0.0027.
\end{figure}

\section{Case IV Different Sample Sizes with backgrounds.}

Next, cases that include backgrounds and allows for difference sample
sizes are presented. Starting again with case 1, it is assumed that
the antineutrino events reduced by a factor $R=2$ and the antineutrino
background is 10 events. \ In this case there are $N_{\mu}=1000$,
$\overline{N}_{\mu}=\frac{N_{\mu}}{2}=500$ and antineutrino background,
$\overline{b}=10$. The CPC averages are $\lambda=51$ and $\overline{\lambda}=\lambda/R+\overline{b}=35.5$.
\ The expected difference (if $\delta_{CP}=-\pi/2$) is $65-R\times\left(37/R+\overline{b}\right)=65-2\times\left(37/2+10\right)=8$.
The expression for this example becomes
\begin{equation}
\sum_{N_{e}=0}^{\infty}\sum_{\overline{N}_{e}=0}^{\infty}\theta\left(N_{e}-2\times\overline{N}_{e}-8+\epsilon\right)\mathscr{P}\left(N_{e},35.5\right)\mathscr{P}\left(\overline{N}_{e},35.5\right)
\end{equation}
and a general expression of the p-value for a null hypothesis test
is,

\begin{equation}
\sum_{N_{e}=0}^{\infty}\sum_{\overline{N}_{e}=0}^{\infty}\theta\left(N_{e}-R\times\overline{N}_{e}-\Delta+\epsilon\right)\mathscr{P}\left(N_{e},\lambda+b\right)\mathscr{P}\left(\overline{N}_{e},\overline{\lambda}/R+\overline{b}\right)
\end{equation}
where $\Delta=N_{e}-R\times\overline{N}_{e}$ is the observed difference,
$R$ is the ratio of $N_{\mu}/\overline{N}_{\mu}$ sample sizes, $\lambda+b$
is the predicted $N_{e}$ null hypothesis events, and $\overline{\lambda}/R+\overline{b}$
is the predicted $\overline{N}_{e}$ null hypothesis events. The relations
between the CP conserving combination of $N_{e}$ and $\overline{N}_{e}$
events, the observed difference $\Delta$, and the p-value region
is presented schematically in Fig. 4. The p-value will be the sum
of the double Poisson terms $\mathscr{P}\left(N_{e},\lambda+b\right)\mathscr{P}\left(\overline{N}_{e},\overline{\lambda}/R+\overline{b}\right)$
in the light red region.

\begin{figure}
\hfill{}\includegraphics[scale=0.9]{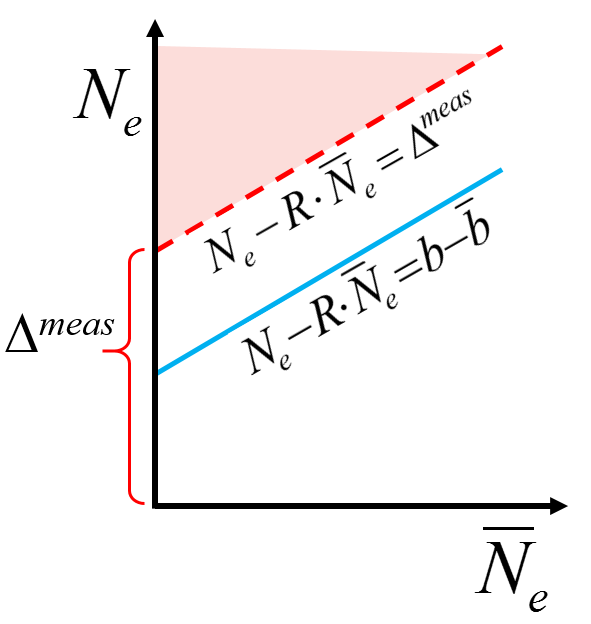}\hfill{}

Fig 4. The CPC events satisfy $N_{e}-R\cdot\overline{N}_{e}=b-\overline{b}$
and will lie on the blue line. The events that have an observed difference
of $N_{e}-R\cdot\overline{N}_{e}=\Delta^{meas}$ will lie on the dashed
red line. The light red solid colored region above the dash red line
is p-value region. The red dashed line p-value boundary has a slope
R and is parallel to the blue line.
\end{figure}

If the ratio $R$ has Gaussian uncertainty, then it can also be varied
to get an averaged p-value. If the predicted number of events have
Gaussian uncertainties, then the Poisson terms are replaced with the
Gaussian smeared Poisson terms. A similar expression for A is, 

\[
\sum_{N_{e}=0}^{\infty}\sum_{\overline{N}_{e}=0}^{\infty}\theta\left(\frac{\left(N_{e}-b\right)-R\times\left(\overline{N}_{e}-\overline{b}\right)}{\left(N_{e}-b\right)+R\times\left(\overline{N}_{e}-\overline{b}\right)}-A\right)
\]

\begin{equation}
\times\mathscr{P}\left(N_{e},\lambda/R+b\right)\mathscr{P}\left(\overline{N}_{e},\lambda/R+\overline{b}\right)
\end{equation}
where $A$ is the measured asymmetry.

\begin{figure}
\hfill{}\includegraphics[scale=0.35]{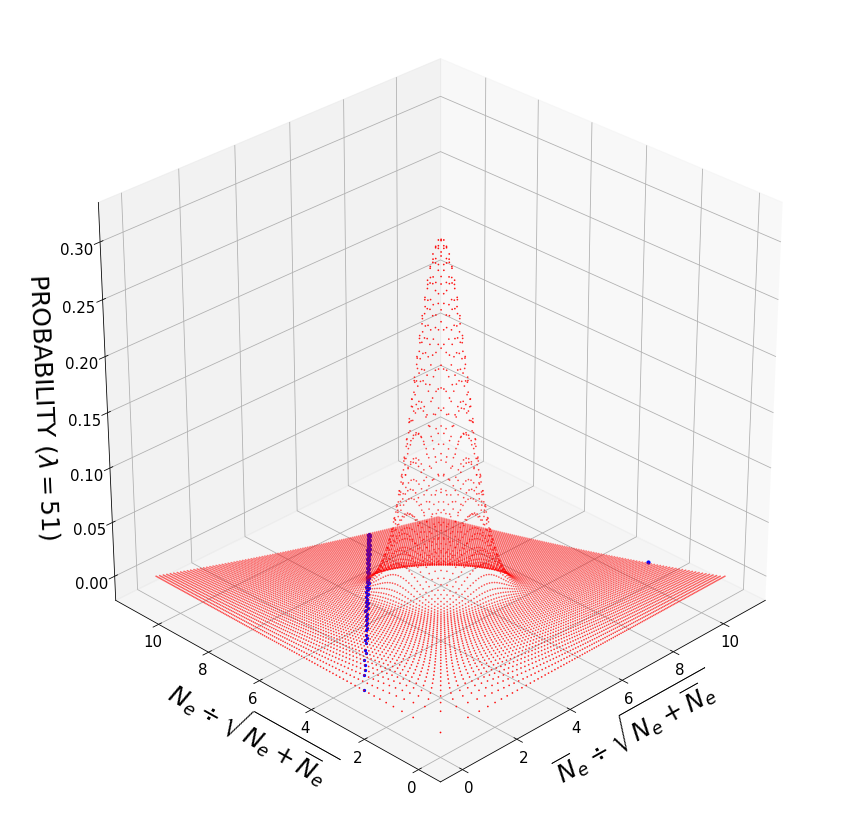}\hfill{}

Fig. 5. The standard deviation distributions of the double Poisson
distributions using the K-T test statistic. The x and y axes are in
standard deviation units of $N_{e}/\sqrt{N_{e}+\overline{N}_{e}}$
and $\overline{N}_{e}/\sqrt{N_{e}+\overline{N}_{e}}$ , respectively.
The Poisson distribution (red dots) has $\lambda=51$ and the blue
boundary corresponds to observed $\Delta=N_{e}-\overline{N}_{e}=65-37=28$
with a p-value of $0.27\%$.
\end{figure}

\section{Case V Unknown Oscillation Probabilities}

In the previous sections, a value of the CPC probability $\lambda$
was assumed to be known and was used in the null hypothesis calculations.
In principle, the average of observed measurements of $N_{e}$ and
$\overline{N}_{e}$ can be used as a statistical estimate for $\lambda$
and instead of using the difference test statistic, $N_{e}-\overline{N}_{e}$,
a better test statistic is, $\frac{N_{e}-\overline{N}_{e}}{\sqrt{N_{e}+\overline{N}_{e}}}$,
where the initial muon neutrino samples are assumed to be equal, $N_{\mu}=\overline{N}_{\mu}$.
This statistic by Krishnamoorthy and Thomson (K-T)\citep{key-8} represents
the difference in units of standard deviations. If the best estimate
of the CPC null hypothesis $\lambda$ is the average of the observed
events, $\left(N_{e}+\overline{N}_{e}\right)/2$, then $\frac{N_{e}-\overline{N}_{e}}{\sqrt{N_{e}+\overline{N}_{e}}}$
is proportional to the difference in units of standard deviations.
In the limit that the Poisson distribution becomes a Normal distribution
it is noted that the p-value boundaries are in units of standard deviations
and the p-value will not depend on a particular value of $\lambda$.

Reconsider Case I, by assuming the observed events are $N_{e}=65$
and $\overline{N}_{e}=37$ and using the test statistic $\frac{N_{e}-\overline{N}_{e}}{\sqrt{N_{e}+\overline{N}_{e}}}$.
The p-value boundary point is at $N_{e}/\sqrt{N_{e}+\overline{N}_{e}}=6.44$
and $\overline{N}_{e}/\sqrt{N_{e}+\overline{N}_{e}}=3.66$ which has
a difference of 2.772. The p-value is now estimated by assuming the
unknown CP conserving probability is given by the estimate $\lambda=\left(N_{e}+\overline{N}_{e}\right)/2$
or 51. 

\begin{equation}
\sum_{N_{e}=0}^{\infty}\sum_{\overline{N}_{e}=0}^{\infty}\theta\left((N_{e}-\overline{N}_{e})/\sqrt{(N_{e}+\overline{N}_{e}}-2.772+\epsilon\right)\mathscr{P}\left(N_{e},51\right)\mathscr{P}\left(\overline{N}_{e},51\right)=0.00271
\end{equation}
The double Poisson is plotted as a function of $N_{e}/\sqrt{N_{e}+\overline{N}_{e}}$
and $\overline{N}_{e}/\sqrt{N_{e}+\overline{N}_{e}}$ with red dots
and the p-value boundary in blue dots that is a 45 degree line that
lies on the boundary point at $N_{e}/\sqrt{N_{e}+\overline{N}_{e}}=6.44$
and $\overline{N}_{e}/\sqrt{N_{e}+\overline{N}_{e}}=3.66$ in Fig.
5. The resulting p-value is $0.271\%$, which is very close to the
results of Case 1 which obtained 0.242\%. This test statistic is largely
insensitive to the true value of $\lambda$. This is readily verified
in our example by recalculating the p-values for different assumed
values of $\lambda$ of 41 or 61 and obtaining the corresponding p-values
of 0.265\% or 0.274\%, respectively. 

This test statistic can be applied to Case II of different sample
sizes. In this case $N_{e}=65$ and $\overline{N}_{e}=37$/2 events
are observed and the p-value boundary point is $N_{e}/\sqrt{N_{e}+4\times\overline{N}_{e}}=5.51$
and $2\times\overline{N}_{e}/\sqrt{N_{e}+4\times\overline{N}_{e}}=3.14$
and the difference is 2.375. The p-value becomes,

\begin{equation}
\sum_{N_{e}=0}^{\infty}\sum_{\overline{N}_{e}=0}^{\infty}\theta\left((N_{e}-2\times\overline{N}_{e})/\sqrt{N_{e}+4\times\overline{N}_{e}}-2.375+\epsilon\right)\mathscr{P}\left(N_{e},51\right)\mathscr{P}\left(\overline{N}_{e},25.5\right)=1.36\%
\end{equation}
The p-value increases to $1.36\%$ compared to the value of $0.9\%$
in Case II. If one assumes values of $\lambda$ of 41 or 61, the resulting
p-values are 1.31\% or 1.41\%, respectively.

Next we apply the statistic to Case III with backgrounds. In this
case, $N_{e}=65$ and $\overline{N}_{e}=37$+10 are observed events
and the p-value boundary is $N_{e}/\sqrt{N_{e}+\overline{N}_{e}}=6.14$
and $\overline{N}_{e}/\sqrt{N_{e}+\overline{N}_{e}}=4.44$. The difference
is 1.70. The p-value becomes,

\begin{equation}
\sum_{N_{e}=0}^{\infty}\sum_{\overline{N}_{e}=0}^{\infty}\theta\left((N_{e}-\overline{N}_{e})/\sqrt{N_{e}+\overline{N}_{e}}-1.70+\epsilon\right)\mathscr{P}\left(N_{e},51\right)\mathscr{P}\left(\overline{N}_{e},51+10\right)=0.41\%
\end{equation}
The p-value slightly increases to $0.41\%$ compared to the value
$0.35\%$ in Case III. If different assumed values of $\lambda$ of
41 or 61 are used, the resulting p-values are slightly changed to
0.31\% or 0.51\%, respectively.

\section{Combining Different Experiments}

As different long baseline neutrino experiments add more data and
perform neutrino and antineutrino oscillation measurements, the results
($ex.$ from experiments A and B) could be combined to obtain a joint
p-value test of the null hypothesis. Assuming both experiments have
a CPC null hypothesis with a common value of $\lambda$ and one-sided
distributions on the same side, then p-values could be extracted from
two 1-dimensional Skellam distributions, $P_{a}^{SK}\left(\Delta_{a}\right)$
and $P_{b}^{SK}\left(\Delta_{b}\right)$. The problem of combining
p-values from two independent measurements has been solved by R. A.
Fisher\citep{key-9} and a simple derivation\citep{key-10} is given
here. Since the p-value of the null hypotheses represents flat distributions
of p-value$_{a}$ and p-value$_{b}$, where each can vary from 0 to
1, a unit square can be formed with axes of these two probabilities.
So unlike the probabilities in Fig. 1(b) and Figs. 2 (a)-(d), the
joint p-value probabilities are flat or constant areas on the p-value$_{a}$
vs p-value$_{b}$ plane as shown in Fig. 6 (a). Suppose experiment
A measures a p-value $p'_{a}$ and experiment B measures a p-value
$p'_{b}$. These two values form a p-value boundary point. Their product
is $c=p'_{a}p'_{b}$ and this forms inside the unit square a boundary
defined by the curve, $p_{b}=c/p_{a}$ as shown in Fig. 6 (a) where
the boundary point lies on the curve defined by $c=p'_{a}p'_{b}$.
The p-value formed by combining these 2 experiments is given by the
region or area inside the unit square corresponding to $p_{a}p_{b}\leq c$
which is below and to the left of this boundary line. This area equals
$-c\ \ln c+c$ and represents the extreme p-value. The area above
and to the right of the boundary curve is 1$-$(p-value). The p-values
whose probabilities correspond to one sided 1, 2 and 3 standard deviations
of normal probabilities are presented in Fig. 6 (a). Suppose there
are two imaginary p-value measurements of 0.4 and 0.0925 from different
experiments. This result appears as a green square in Fig. 6 (a).
These measurements produce the green boundary line whose lower left
area corresponds to 0.159 which is the joint p-value. Since $c=0.4\times0.0925=0.037$,
it is noted that the same combined p-value would be achieved with
two imaginery p-values measurements of $0.37$ and $0.1$. The natural
log of each p-value can be also be plotted as shown in Fig. 6 (b).
The curves become simple diagonal lines and the region above and to
the right of the diagonal lines represents 1$-$(p-value) region.
This can be generalized to n experiments where the probability is
an n dimensional volume and the n-1 dimensional hyper-surface defined
by $c=p_{1}...p_{n}$ which forms the boundary separating the p-value
region and the 1$-$(p-value) region. 

\begin{figure}
\hfill{}\subfloat[]{\includegraphics[scale=0.5]{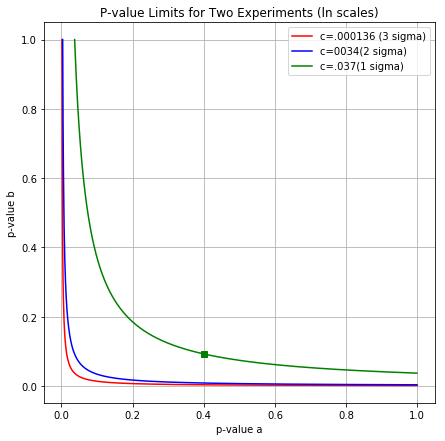}}\subfloat[]{\includegraphics[scale=0.5]{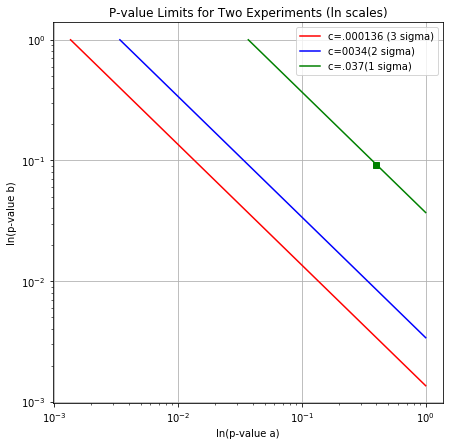}}\hfill{}

Fig. 6. In (a) is the unit square of p-value$_{1}$ versus p-value$_{2}$
with green, blue and red curves representing different probability
boundaries for c = 0.037, 0.0034 and 0.000136, respectively. The areas
below and to the left of each curve is 0.159, 0.0227 and 0.00135,
respectively. If experiment 1 and 2 obtained p-values of .4 and .0925,
respectively, then $c=0.4\times0.0925=0.037$ which corresponds to
the green dot, that lies on the green curve that represents the equation
$p-value_{2}=0.037/p-value_{1}$. The p-value that combines the two
experiments is represented by the area to the left and below the green
curve that equals $-c\ \ln c+c=0.159$ and which is the one sided
probability for 1 sigma normal distribution. The other blue and red
boundaries correspond to the one sided 2 and 3 sigma normal probabilities,
respectively. For small p-values, a more useful graphic is the plot
of ln($p-value_{1}$) vs ln($p-value_{2}$) which forms straight line
diagonal boundaries at 45$^{\circ}$. This is shown in (b), where
the upper left area with respect to these boundaries represents the
1$-$(p-value) for the two experiments.
\end{figure}

\section{Discussion and Summary}

In this paper, the hypothesis of CP conservation that the neutrino
and antineutrino electron appearance oscillation probabilities are
equal, is tested by counting the number of neutrino and antineutrino
events. The inputs to this test static include the observed $N_{e}$
and $\overline{N}_{e}$ and the unoscillated $N_{\mu}$ and $\overline{N}_{\mu}$
, the predicted null hypothesis rates $\lambda+b$ and $\overline{\lambda}+\overline{b}$
, and the ratio R that depends on the neutrino and antineutrino cross
sections, the estimated backgrounds, the reconstruction efficiencies
and relative integrated $\nu$ and $\overline{\nu}$ fluxes. The cases
with equal data samples, with/without backgrounds, unequal data samples,
smeared background rates and CP conserved oscillation probability
rates and unknown CP conserved oscillation rates were discussed. The
p-values can be recalculated as more data is accumulated in experiments
and displayed in modified Skellam distributions or as 3-d plots of
probabilities with p-value boundaries. In addition the p-values from
2 different experiments can be readily combined to produce a joint
p-value to test the null CP conserving hypothesis in neutrino oscillations.
The predictions in this note will be useful to check more sophisticated
and complex statistical tests of CP violation that fit neutrino parameters. 

Finally, methods in this paper can be used to predict the CP conserving
p-value tests assuming a specific set of neutrino mixing parameters
with future scenarios of different amounts of neutrino and antineutrino
beam data. This could be advantageous to optimize the experimental
test for CP violation by varying the amount or mix of neutrino and
antineutrino beam running. 

\section{Acknowledgements}

This work was supported by the U.S. Department of Energy, Office of
Science, under Award Number DE-FOA-0001604. We acknowledge support
from the Program of Research and Scholarly Excellence in High Energy
Physics and Particle Astrophysics at Colorado State University and
we thank Prof. Donald Estep for reading an early draft. 

\section{Appendix}

The appendix describes the dependence between the reconstructed number
of events, backgrounds, oscillation probabilities, cross sections,
neutrino flux and reconstruction efficiencies. This section considers
the case (1) where the neutrino and antineutrino fluxes peak at the
same one energy and the case (2)\ where the neutrino and the antineutrino
fluxes are spread over a range of energies. These relations will require
careful Monte Carlo simulations of the CPC modeling, backgrounds,
detector resolutions and efficiencies and neutrino beam fluxes.

Let's first define the relation of the fluxes and observables. The
number of unoscillated muon neutrinos and antineutrinos in our detectors
are;
\begin{equation}
\begin{array}{c}
\int\Phi^{\prime}\left(E_{\nu_{\mu}}\right)dE_{\nu_{\mu}}=n\left(\nu_{\mu}\right)\\
\int\overline{\Phi}^{\prime}\left(E_{\overline{\nu}_{\mu}}\right)dE_{\overline{\nu}_{\mu}}=n\left(\overline{\nu}_{\mu}\right)
\end{array}
\end{equation}
where the neutrino differential integrated flux is $\Phi^{\prime}\left(E_{\nu}\right)\equiv\partial\Phi\left(E_{\nu}\right)/\partial E_{\nu}$
and the analogous antineutrino barred quantities. The produced number
of oscillated electron appearance neutrinos are;
\begin{equation}
\begin{array}{c}
\int P\left(E_{\nu_{\mu}}\right)\times\Phi^{\prime}\left(E_{\nu_{\mu}}\right)dE_{\nu_{\mu}}=n\left(\nu_{e}\right)\\
\int\overline{P}\left(E_{\overline{\nu}_{\mu}}\right)\times\overline{\Phi}^{\prime}\left(E_{\overline{\nu}_{\mu}}\right)dE_{\overline{\nu}_{\mu}}=n\left(\overline{\nu}_{e}\right)
\end{array}
\end{equation}
where the PMNS probabilities for $P\left(\nu_{\mu}\rightarrow\nu_{e}\right)$
and $\overline{P}\left(\overline{\nu}_{\mu}\rightarrow\overline{\nu}_{e}\right)$
depend on all parameters,
\begin{equation}
\begin{array}{c}
P\left(L/E,\theta_{ij},\Delta m_{ij}^{2},\delta_{CP},MH\right)\\
\overline{P}\left(L/E,\theta_{ij},\Delta m_{ij}^{2},\delta_{CP},MH\right)
\end{array}
\end{equation}
and where MH denotes the mass hierarchy. The true (or produced) and
observed number of appearance $e^{+}/e^{-}$ events at the far detector
are,
\begin{equation}
\begin{array}{c}
\int\sigma\left(E_{\nu}\right)\times P\left(E_{\nu_{\mu}}\right)\times\Phi^{\prime}\left(E_{\nu}\right)dE_{\nu}=n^{true}\left(e^{-}\right)=\frac{n^{obs}\left(e^{-}\right)}{\epsilon_{e}}\\
\int\overline{\sigma}\left(E_{\overline{\nu}}\right)\times\overline{P}\left(E_{\overline{\nu}_{\mu}}\right)\times\overline{\Phi}^{\prime}\left(E_{\overline{\nu}}\right)dE_{\overline{\nu}}=n^{true}\left(e^{+}\right)=\frac{n^{obs}\left(e^{+}\right)}{\epsilon_{\overline{e}}}
\end{array}
\end{equation}
where $\epsilon_{e}$ and $\epsilon_{\overline{e}}$ are the detection
and reconstruction efficiencies and where we use notation $n^{obs}\equiv N^{data}-B^{MC}$.
The number of predicted data events are then,

\begin{equation}
\begin{array}{c}
\lambda+b=\epsilon_{e}\int\sigma\left(E_{\nu}\right)\times P\left(E_{\nu_{\mu}}\right)\times\Phi^{\prime}\left(E_{\nu}\right)dE_{\nu}+B^{MC}\\
\overline{\lambda}+\overline{b}=\epsilon_{\overline{e}}\int\overline{\sigma}\left(E_{\overline{\nu}}\right)\times\overline{P}\left(E_{\overline{\nu}_{\mu}}\right)\times\overline{\Phi}^{\prime}\left(E_{\overline{\nu}}\right)dE_{\overline{\nu}}+\overline{B}^{MC}
\end{array}
\end{equation}
The number of unoscillated muon $\mu^{+}/\mu^{-}$ true (or produced)
events at the near detector extrapolated to the far detector are
\begin{equation}
\begin{array}{c}
\int\sigma\left(E_{\nu}\right)\times\Phi^{\prime}\left(E_{\nu}\right)dE_{\nu}=n^{true}\left(\mu^{-}\right)=\frac{n^{obs}\left(\mu^{-}\right)}{\epsilon_{\mu}}\\
\int\overline{\sigma}\left(E_{\overline{\nu}}\right)\times\overline{\Phi}^{\prime}\left(E_{\overline{\nu}}\right)dE_{\overline{\nu}}=n^{true}\left(\mu^{+}\right)=\frac{n^{obs}\left(\mu^{+}\right)}{\epsilon_{\overline{\mu}}}
\end{array}
\end{equation}
where we assume electron-muon universality, $\sigma_{e}\left(E_{\nu}\right)=\sigma_{\mu}\left(E_{\nu}\right)=\sigma\left(E_{\nu}\right)$
and $\overline{\sigma}_{e}\left(E_{\overline{\nu}}\right)=\overline{\sigma}_{\mu}\left(E_{\overline{\nu}}\right)=\overline{\sigma}\left(E_{\overline{\nu}}\right)$
and the efficiencies are $\epsilon_{\mu}$ and $\epsilon_{\overline{\mu}}$
. The ratio of muon neutrino/antineutrino corrected event rates measured
at the near detector is,

\begin{equation}
r=\frac{\int\overline{\sigma}\left(E_{\overline{\nu}}\right)\overline{\Phi}^{\prime}\left(E_{\overline{\nu}}\right)dE_{\overline{\nu}}}{\int\sigma\left(E_{\nu}\right)\Phi^{\prime}\left(E_{\nu}\right)dE_{\nu}}
\end{equation}
In order to test CP\ violation, we require $P\left(E\right)=\overline{P}\left(E\right)$
for all values of energy $E$.

Case (1) Suppose we have a neutrino and antineutrino beam at the same
one energy given by a Dirac delta function, $\Phi^{\prime}\left(E_{\nu}\right)=\delta\left(E_{\nu}-E_{0}\right)$,
we can readily see
\begin{equation}
\begin{array}{c}
P\left(E_{0}\right)=\frac{\Phi_{e}\left(E_{0}\right)}{\Phi_{\mu}\left(E_{0}\right)}=\frac{\sigma_{e}\left(E_{0}\right)\times P\left(E_{0}\right)\times\Phi_{\mu}\left(E_{0}\right)}{\sigma_{\mu}\left(E_{0}\right)\times\Phi_{\mu}\left(E_{0}\right)}=\frac{n^{true}\left(e^{-}\right)}{n^{true}\left(\mu^{-}\right)}\\
\overline{P}\left(E_{0}\right)=\frac{\overline{\Phi}_{e}\left(E_{0}\right)}{\overline{\Phi}_{\mu}\left(E_{0}\right)}=\frac{\overline{\sigma}_{e}\left(E_{0}\right)\times\overline{P}\left(E_{0}\right)\times\overline{\Phi}_{\mu}\left(E_{0}\right)}{\overline{\sigma}_{\mu}\left(E_{0}\right)\times\overline{\Phi}_{\mu}\left(E_{0}\right)}=\frac{n^{true}\left(e^{+}\right)}{n^{true}\left(\mu^{+}\right)}
\end{array}
\end{equation}
Then $P\left(E_{0}\right)=\overline{P}\left(E_{0}\right)$ can be
tested by using our observed results and checking for inequality of,
\[
\frac{n^{obs}\left(e^{-}\right)/\epsilon_{e}}{n^{obs}\left(\mu^{-}\right)/\epsilon_{\mu}}\neq\frac{n^{obs}\left(e^{+}\right)/\epsilon_{\overline{e}}}{n^{obs}\left(\mu^{+}\right)/\epsilon_{\overline{\mu}}}
\]

\[
\frac{n^{obs}\left(e^{+}\right)/\epsilon_{\overline{e}}}{n^{obs}\left(e^{-}\right)/\epsilon_{e}}\neq\frac{n^{obs}\left(\mu^{+}\right)/\epsilon_{\overline{\mu}}}{n^{obs}\left(\mu^{-}\right)/\epsilon_{\mu}}=r
\]

\begin{equation}
n^{obs}\left(e^{+}\right)-R\times n^{obs}\left(e^{-}\right)\neq0
\end{equation}
where $R=r\times\frac{\epsilon_{\overline{e}}}{\epsilon_{e}}$.

Case (2), Suppose we consider that the flux energy is not a Delta
function and has a flux that is spreadout and smeared as a function
of neutrino/antineutrino energy and they do not have the same shape.
Then we need to be more careful since even if $P=\overline{P}$, is
true at all energies, then it is NOT necessarily true that,
\begin{equation}
\frac{\int\overline{\sigma}\left(E_{\overline{\nu}}\right)\times\overline{P}\left(E_{\overline{\nu}_{\mu}}\right)\times\overline{\Phi}^{\prime}\left(E_{\overline{\nu}}\right)dE_{\overline{\nu}}}{\int\overline{\sigma}\left(E_{\overline{\nu}}\right)\times\overline{\Phi}^{\prime}\left(E_{\overline{\nu}}\right)dE_{\overline{\nu}}}=\frac{\int\sigma\left(E_{\nu}\right)\times P\left(E_{\nu_{\mu}}\right)\times\Phi^{\prime}\left(E_{\nu}\right)dE_{\nu}}{\int\sigma\left(E_{\nu}\right)\times\Phi^{\prime}\left(E_{\nu}\right)dE_{\nu}}
\end{equation}
since the shape of the neutrino and antineutrino flux distribution
as a function of energies are not exactly the same and the cross sections
are not the same.

To allow for this case (2), we calculate the MC probability (assuming
the CP conserving case of $P_{CPC}\left(E\right)=\overline{P}_{CPC}\left(E\right)$
and including MSW effects if necessary) averaged over neutrino and
antineutrino flux energies, so we can add a correction or fudge factor
$\gamma$ which we hope should be very close to unity. Suppose the
MC determines the averaged CPC probabilities,

\begin{equation}
\begin{array}{c}
\left\langle P_{CPC}\left(E_{\nu_{\mu}}\right)\right\rangle =\frac{\int\sigma_{e}\left(E_{\nu}\right)\times P_{CPC}\left(E_{\nu_{\mu}}\right)\times\Phi^{\prime}\left(E_{\nu}\right)dE_{\nu}}{\int\sigma_{\mu}\left(E_{\nu}\right)\times\Phi^{\prime}\left(E_{\nu}\right)dE_{\nu}}\\
\left\langle \overline{P}_{CPC}\left(E_{\overline{\nu}_{\mu}}\right)\right\rangle =\frac{\int\overline{\sigma}_{e}\left(E_{\overline{\nu}}\right)\times P_{CPC}\left(E_{\overline{\nu}_{\mu}}\right)\times\overline{\Phi}^{\prime}\left(E_{\overline{\nu}}\right)dE_{\overline{\nu}}}{\int\overline{\sigma}_{\mu}\left(E_{\overline{\nu}}\right)\times\overline{\Phi}^{\prime}\left(E_{\overline{\nu}}\right)dE_{\overline{\nu}}}
\end{array}
\end{equation}
\begin{equation}
\gamma\equiv\left\langle \overline{P}_{CPC}\left(E_{\overline{\nu}_{\mu}}\right)\right\rangle /\left\langle P_{CPC}\left(E_{\nu_{\mu}}\right)\right\rangle 
\end{equation}
If we now measure or count the four observeables $n^{obs}\left(e^{+}\right)$
and $n^{obs}\left(e^{-}\right)$ at the far detector and $n^{obs}\left(\mu^{+}\right)$
and $n^{obs}\left(\mu^{-}\right)$ at the near detector, the test
of CPV test becomes checking if the following relations are unequal,
\begin{equation}
\gamma\neq\frac{\left(\frac{n^{obs}\left(e^{+}\right)/\epsilon_{\overline{e}}}{n^{obs}\left(\mu^{+}\right)/\epsilon_{\overline{\mu}}}\right)}{\left(\frac{n^{obs}\left(e^{-}\right)/\epsilon_{e}}{n^{obs}\left(\mu^{-}\right)/\epsilon_{\mu}}\right)}
\end{equation}
or
\begin{equation}
\frac{1}{\gamma}\frac{n^{obs}\left(e^{+}\right)/\epsilon_{\overline{e}}}{n^{obs}\left(e^{-}\right)/\epsilon_{e}}\neq\frac{n^{obs}\left(\mu^{+}\right)/\epsilon_{\overline{\mu}}}{n^{obs}\left(\mu^{-}\right)/\epsilon_{\mu}}=r
\end{equation}
or

\begin{equation}
n^{obs}\left(e^{+}\right)-R\times n^{obs}\left(e^{-}\right)\neq0
\end{equation}
where the $R$ from Eq. (10), becomes $R=\gamma\times r\times\frac{\epsilon_{\overline{e}}}{\epsilon_{e}}$.
Note in the final p-value calculations, the R uncertainties should
be included through the gaussian smearing of the Poisson distribution
with the $\sigma\left(\lambda+b\right)$ and $\bar{\sigma}\left(\overline{\lambda}+\overline{b}\right)$
standard deviations and the correlation parameter $\rho$ in Eqn.
\ref{eq:bivariate}.

\end{document}